\documentclass{article}
\usepackage{spconf,amsmath,graphicx}
\usepackage{booktabs}
\usepackage{xcolor}
\usepackage{color}
\usepackage{etoolbox}
\usepackage{tikz}
\usepackage{float}
\usepackage[stable]{footmisc}
\definecolor{skyblue}{RGB}{42, 250, 241}
\definecolor{navy}{RGB}{0, 0, 128}
\definecolor{lightpurple}{RGB}{198, 162, 245}
\definecolor{mypurple}{RGB}{108, 8, 150}
\definecolor{mygreen}{RGB}{35, 219, 87}
\definecolor{darkgray}{RGB}{80, 80, 80}
\usepackage{array}

\newcommand{\baseline}[0]{B1}
\newcommand{\datce}[0]{DAT1-CE}
\newcommand{\mtlce}[0]{MTL1-CE}
\newcommand{\datfocal}[0]{DAT1-F}
\newcommand{\mtlfocal}[0]{MTL1-F}

\newcommand{\datembar}[0]{DAT3}
\newcommand{\mtlembar}[0]{MTL3}
\newcommand{\blc}[0]{B2-labeled(C)}
\newcommand{\bls}[0]{B2-labeled(S)}
\newcommand{\bwc}[0]{B2-w2v(C)}
\newcommand{\bws}[0]{B2-w2v(S)}
\newcommand{\datlc}[0]{DAT2-labeled(C)}
\newcommand{\mtllc}[0]{MTL2-labeled(C)}
\newcommand{\datws}[0]{DAT2-w2v(S)}
\newcommand{\mtlws}[0]{MTL2-w2v(S)}
\newcommand{\datlcar}[0]{DAT3-labeled(C)}
\newcommand{\mtllcar}[0]{MTL3-labeled(C)}
\newcommand{\datwscar}[0]{DAT3-w2v(S)}
\newcommand{\mtlwscar}[0]{MTL3-w2v(S)}

\newrobustcmd*{\mystar}[1]{\tikz{\filldraw[draw=#1,fill=none] star (5,0.5cm);}}

\newrobustcmd*{\mypentagon}[1]{\tikz{\filldraw[draw=#1,fill=none] (0,0.1cm) -- (0.06cm, 0) -- (0.14cm,0) -- (0.2cm, 0.1cm) -- (0.1cm,0.18cm) --(0,0.1cm);}}

\newrobustcmd*{\mysquare}[1]{\tikz{\filldraw[draw=#1,fill=none] (0,0) rectangle (0.2cm,0.2cm);}}

\newrobustcmd*{\mycircle}[1]{\tikz{\filldraw[draw=#1,fill=none] (0,0) circle [radius=0.1cm];}}

\newrobustcmd*{\mytriangle}[1]{\tikz{\filldraw[draw=#1,fill=none] (0,0.2cm) --
(0.2cm,0.2cm) -- (0.1cm,0) -- (0,0.2cm);}}

\usepackage{environ}
\NewEnviron{FitToWidth}[1][\textwidth]{%
\begin{center}
  \makebox[#1]{\resizebox{#1}{!}{\BODY}}%
\end{center}
}

\title{Accent-Robust Automatic Speech Recognition Using Supervised and Unsupervised Wav2vec Embeddings}
\name{\parbox{\linewidth}{Jialu Li$^{\star}$ \thanks{$^{\star}$This work was performed while the author was an intern at Facebook AI.}
 \qquad Vimal Manohar$^{\dagger}$ \qquad Pooja Chitkara$^{\dagger}$ \qquad Andros Tjandra$^{\dagger}$ \qquad Michael Picheny$^{\dagger}$ \\ \centering{\textit{Frank Zhang$^{\dagger}$ \qquad Xiaohui Zhang$^{\dagger}$ \qquad Yatharth Saraf$^{\dagger}$}}}
 }

\address{$^{\star}$Department of ECE \& Beckman Institute, University of Illinois at Urbana-Champaign, IL, USA \\ $^{\dagger}$Facebook AI, USA}
\begin{document}
\ninept
\maketitle

%
\begin{abstract}
Speech recognition models often obtain degraded performance when tested on speech with unseen accents. Domain-adversarial training (DAT) and multi-task learning (MTL) are two common approaches for building accent-robust ASR models. ASR models using accent embeddings is another approach for improving robustness to accents. 
In this study, we perform systematic comparisons of DAT and MTL approaches using a large volume of English accent corpus (4000 hours of US English speech and 1244 hours of 20 non-US-English accents speech). 
We explore embeddings trained under supervised and unsupervised settings: a separate embedding matrix trained using accent labels, and embeddings extracted from a fine-tuned wav2vec model. 
We find that our DAT model trained with supervised embeddings achieves the best performance overall and consistently provides benefits for all testing datasets, and our MTL model trained with wav2vec embeddings are helpful learning accent-invariant features and improving novel/unseen accents.
We also illustrate that wav2vec embeddings have more advantages for building accent-robust ASR when no accent labels are available for training supervised embeddings. 


\end{abstract}

\section{Introduction}
\label{sec:intro}
Automatic speech recognition (ASR) models are expected to be robust to different domains, such as speaker characteristics, accents, dialects, and environmental noise. Nevertheless, ASR models often obtain inferior performance when tested on input from unseen domains. For example, ASR models pretrained on a US English accent don’t generalize well on data from other English accents \cite{jain2018improved, turan2020achieving}. To tackle such problems, transfer learning has drawn a lot of attention in recent studies \cite{cho2018multilingual,kunze2017transfer}.
The key insight of transfer learning is to leverage knowledge learned from a rich-resourced domain and adapt to a lower-resourced domain. Its ultimate goal is to reduce the mismatch of the data distributions between the source data (usually rich-resourced) and target data (usually lower-resourced). In our study, the rich-resourced data is US-English accented speech, and the lower-resourced data is non-US-English accented speech. 

Domain-adversarial training (DAT) \cite{ganin2016domain} has been demonstrated to be beneficial for transfer learning in speech processing \cite{liao2018noise,wang2018unsupervised,sun2018domain}. DAT maps all input domains into a non-discriminative latent space, so the model is able to reuse this latent space for improving performance to unseen input domains. A similar DAT-based approach for learning an accent-invariant representation is to use generative-adversarial nets (GAN) to disentangle accent-invariant and accent-specific acoustic information \cite{chen2020aipnet}. In contrast to DAT, multi-task learning (MTL) with accent recognition as an auxiliary task is another common approach for building accent-robust ASR. 
Many studies discovered that introducing accent embeddings, such as i-vectors \cite{dehak2010front} and x-vectors \cite{snyder2018x} that are known helpful for speaker recognition, as part of input features is another major key to obtain improvements for accented speech \cite{turan2020achieving, viglino2019end,rao2020improved, chen2015improving}. Recently, unsupervised accent modeling has shown to be beneficial for improving accented English speech \cite{redat, 9414833}.
Past studies has investigated the advantages using DAT and MTL on accented speech and unseen accents. \cite{sun2018domain} found that TDNN-based model trained with DAT was consistently superior than MTL on accented Mandarin speech. \cite{redat} showed that RNN-T system, trained on multi-accent English data with DAT, achieved significant improvements on unseen English accents by relabeling one-hot accent labels with unsupervised cluster or soft labels generated from accent classifier. 

Wav2vec (2.0) \cite{schneider2019wav2vec,baevski2020wav2vec} comprises a set of recently proposed models that learn hidden embeddings from massive amounts of untranscribed speech using contrastive loss in an unsupervised learning setting. Its use has resulted in improvement on acoustic modeling for ASR tasks. Several studies also discover that fine-tuned wav2vec embeddings can be beneficial for speech emotion recognition \cite{pepino2021emotion}, speaker verification and language identification \cite{tjandra2021improved} and speech recognition \cite{baevski2021unsupervised}. 

In our experiments, we train a transformer-based CTC \cite{graves2006connectionist} model and demonstrate effectiveness of DAT/MTL learning accent-invariant features. We further quantify the advantages using wav2vec embeddings over labeled embeddings in our ablation study. 
Specifically, this paper makes the following novel contributions: 
\begin{itemize}
  \setlength\itemsep{0.05cm}
    \item We test our models on a large volume of real-world data containing 21 English accents. Note that previous studies usually had less than 10 accents.
    \item We systematically compare both supervised and unsupervised wav2vec embeddings on DAT and MTL models.
    \item We introduce different percentages of corrupted labels for supervised labeled embeddings to illustrate the benefit of using unsupervised wav2vec embeddings. 
\end{itemize}

\section{Data}
\label{sec:data}
Our full dataset consists of about 17k-hours of audio extracted from de-identified public English videos with no personal identifiable information. To protect users' privacy, we cannot make this dataset public. 
We first augment the full dataset using speed perturbation \cite{ko2015audio} and segment it into 10s-long chunks to yield an overall dataset of size 52k hours. In this project, for training data, we selected 10\% of the augmented data to match the original accent distributions of the full dataset. Table \ref{data_distribution} shows the training and testing data distributions for each accent. en\_gb has accents from England, Ireland and Scotland.
All accent labels other than en\_us in our dataset were manually annotated. 
For the testing datasets (``video-types''), en\_us consists of clean (24h), noisy (23h), and extreme (46h) audio; all other accents have the noisy condition only. 

\begin{table}[!h]
\renewcommand{\arraystretch}{1.0}
\setlength{\tabcolsep}{3.0pt}
\centering
\vspace{-0.2cm}
    {
  \begin{FitToWidth}[\columnwidth]
  \begin{tabular}{l|cc||l|cc}
  \toprule
  Accent & Train & Test & Accent & Train & Test \\
  \midrule
  en\_us (US) & 4000 & 92.5 & en\_gb (UK) & 269.8 & 23.2\\
  en\_au (Australia) & 249.1 & 23.3 & en\_in (India) & 142.1 & 16.8\\
    en\_en (England) & 136.7 & 23.2 & en\_ie (Ireland) & 48.0 & 19.9\\
    en\_kr (Korea) & 46.8 & 20.1 & en\_vn (Vietnam) & 42.8 & 15.8\\
    en\_co (Colombia) & 41.6 & 17.6 & en\_ph (Philippine) & 40.0 & 16.2\\
    en\_fr (France) & 34.9 & 25.1 & en\_br (Brazil) & 33.6 & 17.7\\
    en\_mx (Mexico) & 33.0 & 15.4 & en\_ca (Canada) & 31.7 & 19.6\\
    en\_ke (Kenya) & 23.5 & 17.7 & en\_za (South Africa) & 16.8 & 19.0\\
    en\_ng (Nigerian) & 16.7 & 16.3 & en\_eg (Egypt) & 13.1 & 16.9\\
    en\_sq (Scotland) & 13.1 & 9.5 & en\_pk (Pakistan) & 8.4 & 15.7\\
  en\_ar (Argentina) & 2.7 & 9.7 &\\
    \bottomrule
  \end{tabular}
  \end{FitToWidth}
  }
  \vspace{-0.4cm}
  \caption{Training and testing data distributions for accents, in hours. }
  \vspace{-0.3cm}
\label{data_distribution}
\end{table}
For preprocessing data, we extract 80-dimensional log-mel filter energies every 10ms from each utterance. We also encode transcriptions from an inventory of 5000 word pieces, which is the output vocabulary of our ASR model. The word pieces are obtained by training a sentence piece model \cite{kudo2018sentencepiece} on the US English transcripts. 
\section{Model}
\label{sec:model}
\begin{figure}[htb]
\vspace{-0.2cm}
  \centering
  \centerline{\includegraphics[width=0.7\columnwidth]{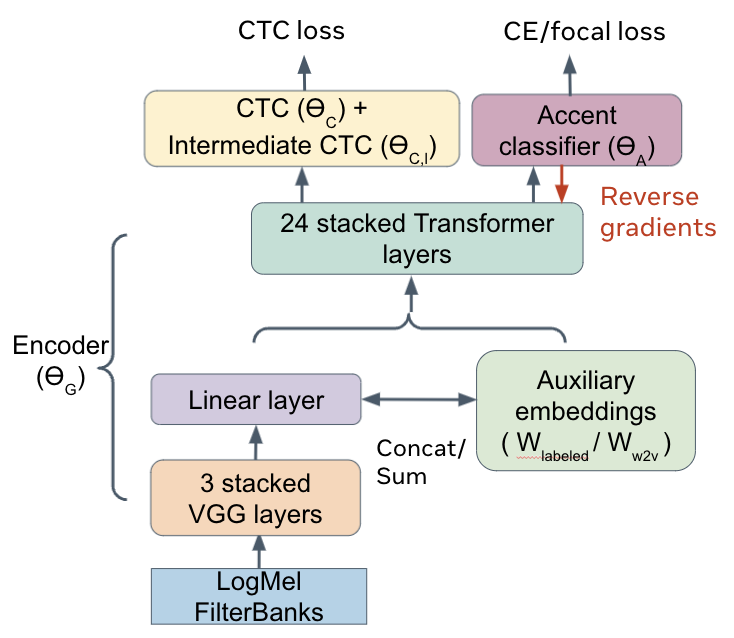}}
\caption{Overall model architecture.}
 \vspace{-0.3cm}
\label{fig:model}
\end{figure}

Figure \ref{fig:model} shows our ASR architecture, a transformer-based CTC architecture with intermediate CTC loss \cite{tjandra2020deja}. The encoder consists of 3 VGG \cite{simonyan2014very} layers and 24 transformer layers. Three VGG layers, each of which has convolutional kernel size 3 and pooling size 2, with increasing number of channels (32, 64, and 128), are used to downsample by a factor of 8 input logmel filterbanks and extract acoustic information. The output of the VGG layers is fed into a linear layer with output dimension 512. Auxiliary embeddings, if used, are either concatenated or weighted summed to the output of linear layer. The combined output is fed into 24 transformer layers to predict word pieces using a CTC loss. Each transformer layer has 8 multi-attention heads with hidden dimension 2048. To enhance ASR performance, we use 3 intermediate CTC losses at the 6$^{\text{th}}$, 12$^{\text{th}}$, and 18$^{\text{th}}$ transformer layers. We place an accent classifier at the 7$^{\text{th}}$ transformer layer, as we empirically observe it performs better than in other positions (layer 13, 19 and 24). Each CTC loss and accent classifier is composed of two layers: one linear layer with output dimension of 256 followed by ReLU activation + one softmax layer. 

We denote $\mathbf{x}$ as the input logmel filterbanks, $\mathbf{y}$ as the word pieces encoded by transcriptions, $\mathbf{y_{a}}$ as the corresponding accent labels, and $N$ as the number of accents. The training objective can be written as Equation (1), where $\mathcal{L}_{CTC}(\mathbf{x,y})$ is the CTC loss, $\mathcal{L}_{CTC,l}(\mathbf{x,y})$ is the CTC intermediate loss, and $\mathcal{L}_{acc}(\mathbf{x,y_{a}})$ is the cross-entropy (CE)/focal loss of accent classifier. We denote $\theta_E$, $\theta_C$, $\theta_{C,l}$ and $\theta_A$ as the parameters of encoder, CTC loss, CTC intermediate loss, and accent classifier respectively. Let $\lambda$ and $\beta$ be weights of intermediate CTC loss and gradient reversal respectively. Equations (2)-(5) show the updated rules used in our experiments, and only the gradients of the accent classifier loss are reversed. Equation (6) specifies the computation for cross-entropy(CE)/focal loss.
{\small
\begin{align}
  \mathcal{L}&=\mathcal{L}_{CTC}(\mathbf{x,y})+\lambda\sum_{l}\mathcal{L}_{CTC,l}(\mathbf{x,y})+\beta\mathcal{L}_{acc}(\mathbf{x,y_{a}})\\
  \theta_E &\xleftarrow[]{} \theta_E - (\frac{\partial \mathcal{L}_{CTC}}{\partial \theta_E} + \lambda\frac{\partial \mathcal{L}_{CTC,l}}{\partial \theta_E} - \beta\frac{\partial \mathcal{L}_{acc}}{\partial \theta_E})\\
  \theta_{C} &\xleftarrow[]{} \theta_{C} - \frac{\partial \mathcal{L}_{CTC}}{\partial \theta_C}\\ 
  \theta_{C,l} &\xleftarrow[]{} \theta_{C,l} - \lambda \frac{\partial \mathcal{L}_{CTC,l}}{\partial \theta_{C,l}} \\
  \theta_{A} &\xleftarrow[]{} \theta_{A} - \beta\frac{\partial \mathcal{L}_{acc}}{\partial \theta_A}\\
\mathcal{L}_{acc} &=
    \begin{cases}
      -\sum_{i=1}^N y_i \log(P(y_i|x_i)) & \text{CE loss}\\
      -\sum_{i=1}^N -(1-P(y_i|x_i))^{\gamma}\log(P(y_i|x_i)) & \text{Focal loss}\\
    \end{cases}   
  \label{eq:dat}
\end{align}
}

In our experiments, we test our data on three models, including a baseline system, DAT and MTL, and we test each model with and without auxiliary embeddings. Our baseline model consists of an encoder, a CTC loss and three CTC intermediate losses without auxiliary embeddings. DAT/MTL has all components of baseline + accent classifier with/without gradient reversal layer respectively.

\vspace{-0.1cm}
\section{Experimental Settings}
We perform two sets of experiments: 1) ``\textit{all}'': this setting trains the model with all 21 accents data and evaluates using 20 non en\_us testing data and all en\_us testing video types; 2) ``\textit{s18}'': this setting trains the model with data from 18 accents, excluding en\_ar (very low-resource), en\_eg (low-resource), and en\_ph (middle-resource), and evaluates using 17 non en\_us accents, 3 novel accents, and all en\_us testing video types. By partitioning our dataset into seen/unseen accents for training and testing, we are able to verify the effectiveness of DAT to learn accent-invariant features and of MTL to learn latent accent representation. For novel accents, we didn't choose all of them to be among the lowest-resourced range to reduce the chance of insufficient test data for learning accent-invariant features.
For all the experiments, we run 100 epochs with 16 GPUs in parallel. We use Adam as our optimizer with an initial learning rate of 0.0012 and forced annealing after epoch 50 by a factor of 0.95. For training the DAT model, we find that setting $\beta>0$ for all training epochs over-trains the accent classifier and fails to learn accent-invariant features. Thus, we set $\beta=0$ for the first 50 epochs to first train the accent classifier until convergence, and then we set $\beta>0$ for the last 50 epochs to back-prop reversed gradients to the encoder. In this way, we observe our accent classifier effectively bring down accent recognition accuracy to around 10\% for the last 50 epochs. For CE loss, we set $\beta=0.03$ and $\lambda=0.3$. For focal loss, we set $\beta=1, \gamma=0.5$, and $\lambda=0.3$. For inference, we use a FST-based decoder with the HLG graph built following the approach in \cite{zhang2020wp_hybrid}. We use a 5-gram language model trained on 14k hours of en\_us transcripts.

\subsection{Supervised labeled embeddings}
\label{sec:labeled_embedding}
We use a separate trainable matrix, $\mathbf{W}_{labeled} \in R^{N \times D}$, as side input, where $N$ is the number of accents in training data and $D$ is the dimension of the hidden embedding. Each row of $\mathbf{W}_{labeled}$ corresponds to an accent, and corresponding rows are either concatenated or weighted and summed to the output of the linear layer. Define $M$ as the hidden dimension of the output of linear layer in the ASR model. We set $D=64$ and $M=448$ for concatenation layer so that the labeled embeddings take a relatively smaller part of the combined features, $D=512$ and $M=512$ for weighted sum layer so that dimensions of the input features and auxiliary embeddings are summable. Similarly, for the wav2vec embeddings (see section \ref{sec:wav2vec_embeddings}), $D=64$ and $M=448$ for the concatenation layer, and $D=768$ and $M=768$ for the weighted sum layer as $D=768$ is the original setting of the wav2vec model. For both types of embeddings, we set the weight as 0.2 for the weighted sum layer.  

\subsection{Unsupervised wav2vec  embeddings}
\label{sec:wav2vec_embeddings}

We use the pretrained wav2vec model from \cite{tjandra2021improved} that consists of 24 transformer layers and learns hidden embeddings from 25 languages of untranscribed speech.
We fine-tune the wav2vec model using our training data by adding an average time pooling layer for producing utterance-level embeddings and a softmax layer for classifying accents. We then extract hidden embeddings from the average time pooling layer as our auxiliary embeddings, $\mathbf{W}_{w2v}$. To achieve the best accent classification accuracy on the validation set, we fine-tune with weighted data samplers by assigning more weight on  en\_us accent data and equal weights across non en\_us accents. For example, we balance our training set with 20\% for en\_us, and 80\% for 20 other non en\_us accents (4\% each).
We achieve the best accent recognition accuracy (for \textit{all}: 74.41\% when D=768 and 75.17\% when D=64; for \textit{s18}: 70.71\% when D=768 and 70.75\% when D=64) when en\_us is assigned 10\% weight and 20 other accents are assigned 90\% (4.5\% each). We confirm that our wav2vec embeddings contain useful accent information based on the visualizations shown in Figure \ref{fig:tsne}. We z-normalize wav2vec embeddings before concatenating or weighted summing it to the output of the linear layer.

\subsection{Visualizations of wav2vec embeddings and accent remap}
\begin{figure}[H]
  \centering
  \vspace{-0.1cm}
  \centerline{\includegraphics[width=9cm]{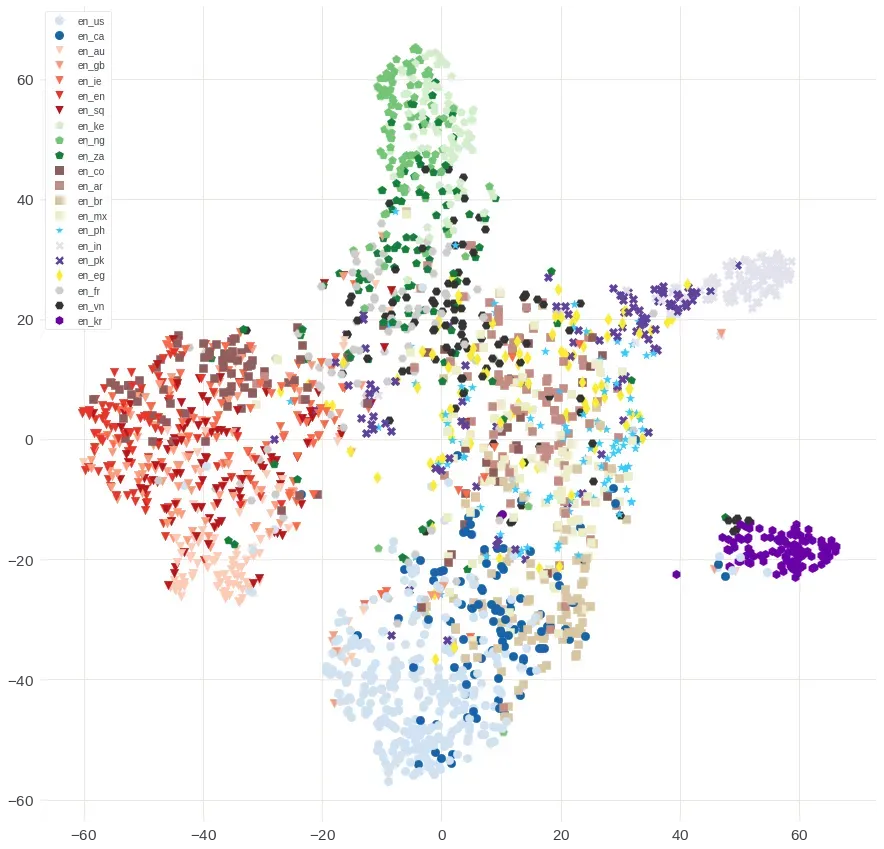}}
\caption{T-SNE plot of wav2vec embeddings. Accents from different regions are labeled with different shapes; accents within the same region are labeled with different ranges of the same color.}
\label{fig:tsne}
  \vspace{-0.3cm}
\end{figure}

To visualize wav2vec embeddings, we randomly select 300 en\_us embeddings and 100 examples from each of the non en\_us embeddings. We first apply Linear Discriminant Analysis to the wav2vec embeddings and reduce the embedding dimension from 768 to 5, and then we apply T-SNE \cite{tsne} with default settings in the sklearn package \cite{scikit-learn} to those reduced-dimension embeddings. Figure \ref{fig:tsne} shows T-SNE plot of embeddings extracted from a wav2vec model pretrained with 8 transformer layers and fine-tuned with all accent labels using CE loss. Based on Figure \ref{fig:tsne}, accents can be grouped by their regions, such as \textcolor{purple}{Europe (en\_au, en\_en, en\_ie, en\_gb, and en\_sq)},  \textcolor{blue}{North America (en\_us and en\_ca)}, \textcolor{mygreen}{Africa (en\_za, en\_ke, and en\_ng)}, \textcolor{brown}{South America (en\_ar, en\_br, en\_mx, and en\_co)}, \textcolor{skyblue}{Southeast Asia (en\_ph)}, \textcolor{lightpurple}{South Asia (en\_in and en\_pk)}, \textcolor{mypurple}{East Asia (en\_kr)}, \textcolor{yellow}{Northeast Africa (en\_eg)}, and \textcolor{darkgray}{Europe/South Asia (en\_fr and en\_vn)}. 
Although en\_fr and en\_vn are not geographically close, 80 years of French colonial rule lead to various loanwords of Vietnamese from French \cite{nguyen2017adaptation}. We thus assign them to one group.
To utilize accent regional information, we remap each accent into larger groups to obtain another set of labels. We observe that en\_co is closer to the European group rather than the South America group, so we assign en\_co into the European group. For \textit{s18} setting, three novel accents are remapped to the closest group (South America) based on Figure \ref{fig:tsne}.

\section{Results}
\label{sec:results}
\vspace{-0.2cm}
\subsection{Baseline results}
\label{sec:baseline_results}
Table \ref{tab:no_emb_baseline_comp} shows performance on the baseline model trained with en\_us data alone versus en\_us + 20 non en\_us accent data.
Training including accented data significantly improves non en\_us word error rates (WER) and slightly benefits en\_us WER. We had observed that previous work (e.g., \cite{turan2020achieving}) typically showed much larger degradation on accented speech when trained only on US-accented speech. We suspected that our en\_us data contained some amount of accented speech. To check this, we ran a wav2vec accent classifier (section \ref{sec:wav2vec_embeddings}), and did find small amounts (2\%) of accented data. This may partially contribute to the smaller than expected performance gap. 
\begin{table}[!h]
  \centering
  \vspace{-0.2cm}
    \setlength{\tabcolsep}{10.0pt}
      \renewcommand{\arraystretch}{0.4}
      \begin{FitToWidth}[\columnwidth]
  \begin{tabular}{c|c|c|c}
  \toprule
  model & training data & non en\_us & en\_us \\
  \midrule
    baseline & en\_us only & 21.7 & 17.7 \\
    \midrule
    baseline & en\_us + non en\_us & 17.0 & 17.3\\
    \bottomrule
  \end{tabular}
  \end{FitToWidth}
  \vspace{-0.5cm}
  \caption{Average WER, in percent (\%), for baseline model without auxiliary embedding trained on en\_us only and en\_us + non en\_us dataset. }
    \vspace{-0.5cm}
  \label{tab:no_emb_baseline_comp}
\end{table}

\subsection{Embeddings effects on baseline, DAT and MTL models}
Table \ref{tab:no_emb} summarizes important results for experiments without auxiliary embeddings. Models trained with focal loss ({\datfocal } and \mtlfocal) achieve the best results on en\_us testing data and comparable results on non en\_us data in both \textit{all} and \textit{s18} settings, compared with CE loss ({\datce }  and \mtlce). {\mtlce } shows worse results than the baseline (\baseline). Thus, we train the rest of the experiments using focal loss.

\begin{table}[ht]
  \centering
  \vspace{-0.2cm}
  \setlength{\tabcolsep}{8.0pt}
  \renewcommand{\arraystretch}{0.4}
  \begin{FitToWidth}[\columnwidth]
  \begin{tabular}{l|c|cc|c|cc}
  \toprule
  model &accent & \multicolumn{2}{c|}{ non en\_us} & \multicolumn{1}{c|}{ novel} & \multicolumn{2}{c} { en\_us }\\
  &loss &\textit{all}  & \textit{s18} & \textit{s18} & \textit{all}  & \textit{s18} \\
  \midrule
    \textbf{\baseline}: baseline& - & 17.0 & 16.5 & 21.2 & 17.3 & 17.1\\
    \midrule
    \textbf{\datce} & CE & 16.9 & 16.3 & 21.1 & 17.2 & 17.1\\
    \textbf{\mtlce} &  & 17.1 & 17.0 & 21.1 & 17.3& 17.2\\
    \midrule
    \textbf{\datfocal} & focal & {17.0} & 16.4 & 21.3 & \textbf{17.0} & \textbf{17.0} \\
    \textbf{\mtlfocal} & & 16.9 & 16.4 &  21.2 & 17.1 & \textbf{17.0} \\
    \bottomrule
  \end{tabular}
  \end{FitToWidth}
      \vspace{-0.4cm}
  \caption{Average WER for baseline, DAT and MTL without embeddings. Models are named with initials+numbers+accent classifier loss settings, e.g, {\datce } means DAT trained with CE loss. The best result for each setting is \textbf{bolded} across Table \ref{tab:no_emb} and \ref{tab:emb}.}
  \label{tab:no_emb}
\end{table}

Table \ref{tab:emb} presents results of baseline, DAT and MTL models using embeddings. 
We observe that {\blc } is generally better than {\bls }, while {\bws } is generally better than {\bwc }, which can be due to the mismatch distributions between the input features and wav2vec embeddings. Thus, we keep this setting for the rest of the experiments.
{\datlc }, our overall best model, shows equal or superior performances than {\mtllc } and {\blc } for all testing datasets, which demonstrates the effectiveness of DAT learning accent-invariant information. 
We find that {\mtllc } is not better than {\blc }, which suggests that the accent classifier is not essential when we feed ground-truth labeled embeddings as part of the input. However, accent classifier is helpful when we feed wav2vec embeddings, as {\mtlws } benefits novel accents the most than other models using wav2vec embeddings.
For experiments using the accent remap technique, regional information is mainly beneficial for {\datwscar } and {\mtlwscar } trained on non en\_us and en\_us datasets, compared to corresponding settings without accent remap ({\datws } and \mtlws). 
A typical trend of our results in Table \ref{tab:emb} is that we have slightly worse performances on en\_us and relatively small improvements on non en\_us and novel accents compared to our baseline. This may be explained by 1) our en\_us training data is contaminated by a small amount of other accents (see section \ref{sec:baseline_results}), and 2) comparing against a baseline model trained on multi-accent data yields smaller improvements as opposed to comparing against a single yet mismatched accent model. 
Similar trends can be found in \cite{redat}: DAT yields smaller improvements on unseen en-AU accent data over the ``data pooling'' baseline compared to accent-specific models with individual accent embeddings.

\begin{table}[ht]
    \vspace{-0.2cm}
  \centering
  \setlength{\tabcolsep}{8.0pt}
  \renewcommand{\arraystretch}{0.4}
  \begin{FitToWidth}[\columnwidth]
  \begin{tabular}{l|cc|c|cc}
  \toprule
  model & \multicolumn{2}{c|}{non en\_us} & \multicolumn{1}{c|}{novel} & \multicolumn{2}{c} {en\_us}\\
   &\textit{all}  & \textit{s18} & \textit{s18} & \textit{all}  & \textit{s18}\\
  \midrule
    \textbf{\baseline}: baseline & 17.0 & 16.5 & 21.2 & 17.3 & 17.1\\
    \midrule\midrule
    \textbf{\blc} & \textbf{16.4} & \textbf{15.9} & \textbf{20.9} & 17.3 & 17.3\\
    \textbf{\bls} & 16.5 & 16.1 & 21.4 & \textbf{17.0} & 17.3 \\
    \midrule
    \textbf{\bwc} & 17.0 & 16.4 & 21.2 & 17.1 & 17.4\\
    \textbf{\bws} &16.7 & 16.4 & 21.1 & 17.3 & 17.3 \\
    \midrule
    \textbf{\datlc}& \textbf{16.4} & \textbf{15.9} & \textbf{20.9} & {17.1} & {17.2} \\
    \textbf{\mtllc} & 16.5 & 16.7 & 21.0 & {17.2} & 17.4 \\
    \midrule
    \textbf{\datws} & 16.9 & 16.4 & {21.1} & 17.3 & 17.6 \\
    \textbf{\mtlws} & 16.9 & {16.3} & {21.0} & {17.1} & 17.5 \\ 
    \midrule \midrule
    \textbf{\datlcar} & 16.8 & 16.4 & 21.1 & 17.4 & 17.5 \\
    \textbf{\mtllcar} & 16.7 & 16.3 & 21.0 & \textbf{{17.0}} & {17.1} \\
    \midrule
    \textbf{\datwscar} & 16.8 & {16.4} & {21.1} & {17.1} & {17.2} \\
    \textbf{\mtlwscar} & 16.8 & 16.4 & 21.3 & {17.1} & {17.1} \\
    \bottomrule
  \end{tabular}
  \end{FitToWidth}
    \vspace{-0.4cm}
  \caption{Average WER for baseline, DAT and MTL with labeled/wav2vec embeddings using focal loss. C means concatenation layer; S means weighted sum layer, as described in Section \ref{sec:model}. Models are named with initials+numbers+embedding, e.g, {\datlc } means DAT trained with labeled embeddings in concatenation layer (C). Groups of {\datembar } and {\mtlembar } are experiments using accent remap technique. The best result for each setting is \textbf{bolded} across Table \ref{tab:no_emb} and \ref{tab:emb}. 
  }
  \label{tab:emb}
    \vspace{-0.3cm}
\end{table}

\subsection{Ablation study of corrupted labels on labeled embeddings}
Although we achieve the best performance with supervised labeled embeddings, we may not always have accurate accent labels available. To understand the effects of accurate labels, we perform an ablation study by introducing different amounts of corrupted labels for labeled embeddings during inference. For generating corrupted labels, we use random and incorrect labels replacing part of the ground-truth labels for non en\_us and en\_us datasets in both \textit{all} and \textit{s18} settings. 
Because ground-truth labels are not defined for novel accents in \textit{s18}, we test two types of embeddings for its ground-truth: 1) untrained embeddings that are randomly initialized and 2) en\_us embeddings. Table \ref{tab:corrupted_label} shows relevant results. With more corrupted labels, WER increases across three testing datasets as expected. We can estimate that if 10-25\% of the labels are corrupted, wav2vec embeddings will have superior performances than labeled embeddings. For novel accents, untrained embeddings are more helpful than en\_us embeddings even though en\_us embeddings are better trained.   

\begin{table}[ht]
\vspace{-0.2cm}
  \centering
  \setlength{\tabcolsep}{8.0pt}
  \renewcommand{\arraystretch}{0.4}
  \begin{FitToWidth}[\columnwidth]
  \begin{tabular}{c|cc|cc|cc}
  \toprule
   & \multicolumn{2}{c|}{non en\_us } & \multicolumn{2}{c|}{novel } & \multicolumn{2}{c} {en\_us}\\
   & \textit{all}  & \textit{s18} & \textit{s18}*  & $\textit{s18}^{\dagger}$ & \textit{all}  & \textit{s18} \\
  \midrule
    0\% & {16.4} & {15.9} & {20.9} & 21.6 & 17.3 & 17.3\\
    \midrule
    10\% & 16.6 & 16.2 & 20.9 & 21.6& 17.3 & 17.2 \\
    25\% & 16.9  & 16.5 & 21.1 & 21.6 & 17.4& 17.3 \\
    50\% & 17.4 & 16.9  & 21.3 & 21.6 & 17.4 & 17.3 \\
    \midrule
    \textbf{\bws} &16.7 & 16.4 & 21.1 & - & 17.3 & 17.3 \\
    \bottomrule
  \end{tabular}
  \end{FitToWidth}
      \vspace{-0.5cm}
  \caption{Average WER for \blc \space trained with different percentage of corrupted labels. Novel accents labels  were tested for both untrained(*) and en\_us($\dagger$) embeddings. Results for \bws \space are copied to the last row for easy comparisons.}
  \label{tab:corrupted_label}
\end{table}

\vspace{-0.5cm}
\section{Conclusion}
\label{sec:conclusion}
We investigate three transformer-based CTC models: baseline, DAT and MTL, and we test each model with two types of embeddings, supervised and unsupervised wav2vec embeddings, on a dataset consisting of en\_us accent and 20 non en\_us accents. We find that DAT trained with labeled embeddings encourages the model to learn accent-invariant features and provides benefits across all testing datasets. MTL trained with wav2vec embeddings performs the best for novel accents among all other corresponding models. We show that using wav2vec embeddings is beneficial when no accent labels are available for training labeled embeddings, and results from our ablation study illustrate that wav2vec embeddings are better than labeled embeddings when 10-25\% accent training labels are corrupted.




\bibliographystyle{IEEEbib}
\bibliography{strings,refs}

\end{document}